\documentclass[%
 preprint,
 amsmath,amssymb,
 aps,
 pre,
showkeys
]{revtex4-2}

\usepackage{graphicx}
\usepackage{dcolumn}
\usepackage{bm}
\usepackage{hyperref}
\usepackage{xcolor}
\usepackage{ctable}

\begin{document}

\title{Transition temperature and thermodynamic properties of homogeneous weakly interacting Bose gas in self-consistent Popov approximation}

\author{Nguyen Van Thu$^*$, Pham Duy Thanh and Lo Thi Thuy}
\affiliation{Department of Physics, Hanoi Pedagogical University 2, Hanoi, Vietnam}
\email[]{nvthu@live.com}


\date{\today}

\begin{abstract}
This study utilizes the Cornwall-Jackiw-Tomboulis effective action approach combined with variational perturbation theory to investigate the relative shift in the transition temperature of a homogeneous, repulsive, weakly interacting Bose gas compared to that of an ideal Bose gas. By applying both the one-loop and self-consistent Popov approximations, the universal form of the relative shift in the transition temperature is derived, demonstrating its proportionality to the $s$-wave scattering length. The results exhibit excellent agreement with those obtained from precise Monte Carlo simulations. Furthermore, the zero-point energy and various thermodynamic properties are examined in both the condensed and normal phases. A comparison with experimental data reveals an excellent agreement, further validating the findings.
\end{abstract}

\keywords{Interacting Bose gas,self-consistent Popov approximation, transition temperature, thermodynamic properties}

\maketitle

\section{Introduction\label{sec1}}

The Bose-Einstein condensation (BEC) of a homogeneous repulsive weakly interacting Bose gases remains a highly topical and promising area of research, despite having been theoretically predicted nearly a century ago \cite{Bose1924,Einstein1924}. It was pointed out that a system of bosonic atoms undergo  a condensate phase transition as its temperature is lowered to the critical temperature. Studies of BEC truly blossomed after the condensation of Bose gas was observed experimentally \cite{Anderson1995} and have yielded numerous significant results. For a perfect Bose gas, the transition temperature is \cite{Huang1987,Pethick2008},
\begin{eqnarray}
T_C^{(0)}=\frac{2\pi\hbar^2}{mk_B}\left[\frac{\rho}{\zeta(3/2)}\right]^{2/3},\label{Tideal}
\end{eqnarray}
in which $\hbar$ and $k_B$ are reduced Planck and Boltzmann constants, $m$ is the atomic mass and $\rho$ is the particle density. The zeta function is defined as $\zeta(x)=\sum_{n=1}^\infty 1/n^x$.

In a system of imperfect Bose gas, the bosonic atoms interact with one another. Various potential models have been proposed to describe these interactions, including two-body and three-body potentials \cite{Wu1959}; hard-sphere, soft-sphere, and hard-core square-well models \cite{Giorgini1999} as well as pseudo- and exponential potentials \cite{Carlen2021}. From these studies, it has been concluded that the interatomic interaction in a dilute Bose gas is characterized by the $s$-wave scattering length
$a_s$. This interaction modifies the transition temperature of the interacting Bose gas compared to that of the ideal Bose gas. To leading order in the $s$-wave scattering length, the relative shift in the transition temperature of a homogeneous, repulsive, weakly interacting Bose gas, compared to that of an ideal Bose gas, follows a universal form
\begin{eqnarray}
\frac{\Delta T_C}{T_C^{(0)}}\approx c.\alpha_s^a,\label{shift0}
\end{eqnarray}
where the gas parameter $\alpha_s=\rho a_s^3$ is defined in terms of $\rho$, the particle number density and the scattering length $a_s$. The determination of the constants $a$ and $c$ has been the subject of prolonged debate and controversy. Most studies suggest $a=1/3$ \cite{Toyoda1982,Grueter1997,Arnold2001,Napiorkowski2017} (i.e. the shift is proportional to $a_s$ and $\rho^{1/3}$), though some results indicate $a=1/6$ \cite{Kleinert2005,Huang1999,Seiringer2009} (i.e. the shift is proportional to $a_s^{1/2}$ and $\rho^{1/6}$). Similarly, the value of the constant $c$ has been contentious: while some calculations yield negative values \cite{Toyoda1982,Wilkens2000}, the majority report positive values over a broad range \cite{Grueter1997,Arnold2001,Davis2003}. A remarkably high value of $c=3.059$ was reported in Ref. \cite{SouzaCruz2001}, while an even greater value of $c=4.66$ was documented in Ref. \cite{Stoof1992}. Furthermore, experimental data have indicated a significant value of $c=5.1\pm0.9$ in Ref. \cite{Reppy2000}.

Most recent investigations \cite{AlSugheir2023} have suggested that, at low densities, the relative shift exhibits a dependence on $\rho^{2/3}$, whereas alternative analyses \cite{Vianello2024} propose a scaling with $\rho^{1/6}$ and $a_s^{1/2}$.  The numerical computation results are not consistent. Specifically, Gr\"{u}ter {\it et.al.} obtained $c=0.34\pm 0.06$ in Ref. \cite{Grueter1997}, whereas the authors of Refs. \cite{Kashurnikov2001,Arnold2001a} reported $c=1.29\pm 0.05$.

To study the dilute Bose gas at finite temperatures, most authors have utilized the Hartree-Fock-Bogoliubov (HFB) theory, as documented in the literature \cite{Pethick2008,Griffin2009}. However, the HFB theory presents certain disadvantages, including issues related to conservation laws and the presence of a gap in the energy spectrum of excitations, which arise due to the anomalous density, as discussed in Ref. \cite{Griffin1996}.
To address these challenges, several approximations have been proposed, such as the Popov, Beliaev-Popov, and $T$-matrix approximations \cite{Andersen2004,Shi1998}, with the Popov approximation being particularly prevalent. Nevertheless, the Popov approximation results in a first-order phase transition in the condensation phase transition, which is an unphysical outcome. To overcome this inconsistency, Yakulov and collaborators \cite{Yukalov2006} introduced a framework involving Lagrange multipliers with two chemical potentials. In their approach, one chemical potential primarily resolves the conservation problem, while the other ensures the energy spectrum remains gapless. Despite this refinement, their results indicated that the transition temperature for a weakly interacting Bose gas coincides with that of an ideal Bose gas. A similar issue was encountered when Haugset {\it et. al.}  \cite{Haugset1998}  applied the Cornwall–Jackiw–Tomboulis (CJT) effective action approach in the one-loop approximation to investigate the dilute Bose gas. Fortunately, this conclusion was subsequently disproven and addressed by Kleinert {\it et. al.} \cite{Kleinert2005} through the application of variational perturbation theory. While the relative shift in the transition temperature was found to deviate from zero, the authors clarified that their result, characterized by a square-root behavior, was incorrect. In the present study, we integrate the CJT effective action approach within the one-loop approximation with variational perturbation theory in the Popov approximation. This combination aims to refine and extend the findings in Refs. \cite{Kleinert2005,Haugset1998}, particularly concerning the relative shift in transition temperature and the thermodynamic quantities of a homogeneous repulsive weakly interacting Bose gas. Indeed, thermodynamic properties of the interacting Bose gas are investigated in both condensed and normal phases.

This paper is organised as follows. In Section \ref{sec:2} focus on finding the relative shift of transition temperature with respect to the ideal Bose gas by means of the CJT effective action approach at finite temperature. Section \ref{sec:3} devotes for thermodynamic quantities of a homogeneous dilute weakly interacting Bose gas . Finally, we present the conclusion and future outlook in Section \ref{sec:4}.

\section{Transition temperature in the self-consistent Popov approximation\label{sec:2}}

\subsection{Self-consistent Popov approximation}

In this Section, we will investigate effect of repulsive weakly interatomic interaction to transition temperature of the Bose gas by using CJT effective action in the one-loop approximation. In addition, we investigate it in the Popov approximation with the self-consistency as pointed out in Ref. \cite{Kleinert2005}. To do this, we start with a homogeneous dilute Bose gas described by the Lagrangian density \cite{Pethick2008},
\begin{equation}
{\cal L}=\psi^*\left(-i\hbar\frac{\partial}{\partial t}-\frac{\hbar^2}{2m}\nabla^2\right)\psi-\mu\left|\psi\right|^2+\frac{g}{2}\left|\psi\right|^4,\label{eq:1}
\end{equation}
wherein $\mu$ is the chemical potential. The field operator $\psi(\vec{r},t)$ depends on both the coordinate $\vec{r}$ and time $t$. The interatomic interaction potential between the atoms can be chosen as the hard-sphere model. In the Born approximation, the strength of the interaction between pairwise atoms is determined via the $s$-wave scattering length $a_s$ as $g=4\pi\hbar^2a_s/m$. Now, thermodynamic stability requires that $g>0$, i.e., the boson interactions are repulsive \cite{Pethick2008}. In the tree-approximation, the expectation value $\psi_0$ of the field operator is independent of both coordinate and time. Therefore, the GP potential can be read off from the Lagrangian density (\ref{eq:1}),
\begin{equation}
V_{\rm GP}=-\mu\left|\psi_0\right|^2+\frac{g}{2}\left|\psi_0\right|^4.\label{eq:VGP}
\end{equation}
Without any external fields and macroscopic part of the condensate moving as a whole, the lowest energy solution $\psi_0$ is real and plays the role of the order parameter. Minimizing the potential \eqref{eq:VGP} with respect to the order parameter, one arrives at the gap equation
\begin{eqnarray}
\psi_0(-\mu+g\left|\psi_0\right|^2)=0.\label{gaptree}
\end{eqnarray}
The square of the order parameter $\rho_0\equiv\left|\psi_0\right|^2$ is defined as the density of the condensate. Eq. (\ref{gaptree}) yields the condensed density in the condensed phase
\begin{equation}
\rho_0=\frac{\mu}{g}.\label{eq:psi0}
\end{equation}
Paying attention to Eq. (\ref{gaptree}), we can write the propagator in tree-level as follows
\begin{equation}
D_0(k)=\frac{1}{\omega_n^2+E_{\rm{(tree)}}^2(k)}\left(
              \begin{array}{cc}
                \varepsilon_k & \omega_n \\
                -\omega_n &  \varepsilon_k+2\mu\\
              \end{array}
            \right),\label{eq:protree}
\end{equation}
where $\varepsilon_k=\hbar^2k^2/2m$ is a function of wave vector $\vec{k}$, the $n$th Matsubara frequency for bosons is defined as $\omega_n=2\pi n/\beta$ with $n\in{\mathbb{Z}}$ and $\beta=1/k_BT$. Examining poles of the propagator (\ref{eq:protree}) gives us the dispersion relation in tree-level
\begin{equation}
E_{\rm{(tree)}}(k)=\sqrt{\varepsilon_k\left(\varepsilon_k+2\mu\right)}.\label{dispertree}
\end{equation}
It is obvious that the energy spectrum of the excitations is gapless, which associates with the spontaneous breaking of the $U(1)$ continuous symmetry. This implies that our system obeys the Goldstone theorem \cite{Goldstone1961}. This fact is referred as the Hugenholtz-Pines theorem \cite{Hugenholtz1959} at zero temperature and a more general proof for all value of temperature was given by Hohenberg  and Martin \cite{Hohenberg1965}.

Beyond the mean field theory, the field operator is decomposed in form \cite{Andersen2004}
\begin{equation}
\psi\rightarrow \psi_0+\frac{1}{\sqrt{2}}(\psi_1+i\psi_2),\label{eq:shift}
\end{equation}
in which $\psi_1$ and $\psi_2$ are associated with fluctuations of the field. Inserting (\ref{eq:shift}) into (\ref{eq:1}) one obtains the interacting Lagrangian density \cite{Thu2024}
\begin{eqnarray}
{\cal L}_{\rm int}=\frac{g}{2}\psi_0\psi_1(\psi_1^2+\psi_2^2)+\frac{g}{8}(\psi_1^2+\psi_2^2)^2.\label{eq:Lint}
\end{eqnarray}
In the one-loop approximation, the CJT effective potential per unit volume can be read from (\ref{eq:Lint}) as follows \cite{VanThu2017},
\begin{eqnarray}
V_\beta=-\mu\left|\psi_0\right|^2+\frac{g}{2}\left|\psi_0\right|^4+\frac{1}{2}\int_\beta{\rm{tr}}\ln D^{-1}(k),\label{Vbeta}
\end{eqnarray}
in which the notation $\int_\beta$ is abbreviated for
\begin{eqnarray*}
\int_\beta f(k)=\frac{1}{\beta}\sum_{n=-\infty}^{+\infty} \int \frac{d^3\vec{k}}{(2\pi)^3}f(k,\omega_n).
\end{eqnarray*}
The propagator in the one-loop approximation $D(k)$ is deduced from (\ref{Vbeta})
\begin{equation}
D(k)=\frac{1}{\omega_n^2+E^2(k)}\left(
              \begin{array}{cc}
                \varepsilon_k-\mu+g\left|\psi_0\right|^2 & \omega_n \\
                -\omega_n &  \varepsilon_k-\mu+3g\left|\psi_0\right|^2\\
              \end{array}
            \right).\label{eq:protree}
\end{equation}
Analogously, the dispersion relation in the one-loop approximation is
\begin{eqnarray}
E(k)=\sqrt{\left[\varepsilon_k-\mu+2g\left|\psi_0\right|^2\right]^2-g^2\left|\psi_0\right|^4}.\label{Eone}
\end{eqnarray}
To proceed further we employ the rule for the summation of Matsubara frequencies \cite{Schmitt2010},
\begin{equation}
\frac{1}{\beta}\sum_{n=-\infty}^{n=+\infty}\ln\left[\omega_n^2+E^2(k)\right]=E(k)+\frac{2}{\beta}\ln\left[1-e^{-\beta E(k)}\right].\label{rule}
\end{equation}
Applying (\ref{rule}) to the last term in right-hand side of (\ref{Vbeta}) the CJT effective potential can be rewritten as
\begin{eqnarray}
V_\beta=&&-\mu\left|\psi_0\right|^2+\frac{g}{2}\left|\psi_0\right|^4+\frac{1}{2}\int_\beta \frac{d^3\vec{k}}{(2\pi)^3}E(k)\nonumber\\
&&+\frac{1}{\beta}\int \frac{d^3\vec{k}}{(2\pi)^3}\ln\left[1-e^{-\beta E(k)}\right],\label{Vbeta1}
\end{eqnarray}
By extremizing the CJT effective potential (\ref{Vbeta1}) with respect to $\psi_0$ one arrives at the gap equation
\begin{eqnarray}
&&-\mu+g\left|\psi_0\right|^2+\frac{1}{2}\int\frac{d^3\vec{k}}{(2\pi)^3}\frac{2g\left[\varepsilon_k-\mu+2g\left|\psi_0\right|^2\right]-g^2\left|\psi_0\right|^2}{E(k)}\nonumber\\
&&+\int\frac{d^3\vec{k}}{(2\pi)^3}\frac{2g\left[\varepsilon_k-\mu+2g\left|\psi_0\right|^2\right]-g^2\left|\psi_0\right|^2}{E(k)\left[e^{\beta E(k)}-1\right]}=0.\label{gapone}
\end{eqnarray}
Combining Eqs. (\ref{gapone}) and (\ref{eq:psi0}) yields a relation the condensed density \cite{Kleinert2005}
\begin{eqnarray}
\rho_0=\frac{\mu}{g}-\int\frac{d^3\vec{k}}{(2\pi)^3}\frac{2\varepsilon_k+\mu}{\sqrt{\varepsilon_k^2+2\mu\varepsilon_k}}\left[\frac{1}{2}+\frac{1}{e^{\beta\sqrt{\varepsilon_k^2+2\mu\varepsilon_k}}-1}\right].\label{rho0}
\end{eqnarray}
We next calculate the pressure of the system, which is defined as the minus of the CJT effective potential (\ref{Vbeta1}) taking at its minimum, i.e., satisfying the condition (\ref{rho0}). It is easy to find
\begin{eqnarray}
{\cal P}=\frac{\mu^2}{2g}-\frac{1}{2}\int\frac{d^3\vec{k}}{(2\pi)^3}\sqrt{\varepsilon_k^2+2\mu\varepsilon_k}-\frac{1}{\beta}\int\frac{d^3\vec{k}}{(2\pi)^3}\ln\left[1-e^{-\beta\sqrt{\varepsilon_k^2+2\mu\varepsilon_k}}\right].\label{pressure}
\end{eqnarray}
The total particle density is defined as derivative of the pressure with respect to the chemical potential
\begin{eqnarray}
\rho=\frac{\partial {\cal P}}{\partial\mu}.\label{density}
\end{eqnarray}
Plugging (\ref{pressure}) into (\ref{density})  yields
\begin{eqnarray}
\rho=\frac{\mu}{g}-\int\frac{d^3\vec{k}}{(2\pi)^3}\frac{\varepsilon_k}{\sqrt{\varepsilon_k^2+2\mu\varepsilon_k}}\left[\frac{1}{2}+\frac{1}{e^{\beta\sqrt{\varepsilon_k^2+2\mu\varepsilon_k}}-1}\right].\label{rho}
\end{eqnarray}
In relevant experiments, the measured quantity is the non-condensed density \cite{Thu2023}.  Eliminating the chemical potential from (\ref{rho0}) and substituting into (\ref{rho}) one attains the non-condensed density
\begin{eqnarray}
\rho-\rho_0=\int\frac{d^3\vec{k}}{(2\pi)^3}\frac{\varepsilon_k+\mu}{\sqrt{\varepsilon_k^2+2\mu\varepsilon_k}}\left[\frac{1}{2}+\frac{1}{e^{\beta\sqrt{\varepsilon_k^2+2\mu\varepsilon_k}}-1}\right].\label{rho1}
\end{eqnarray}
Eq. (\ref{rho1}) represents the non-condensed density in the so-called Popov approximation \cite{Andersen2004,Shi1998}.

To ensure self-consistency, the variational perturbation method introduced by Kleinert {\it et.al.} \cite{Kleinert2005} is employed. This method involves introducing a variational parameter $M$ defined by the relation $\mu=M+r\eta$ where $\eta$ is a counting loop expansion parameter and $r=(\mu-M)/\eta$. This parameter results in a trial CJT effective potential $V_\beta^{\rm(trial)}$, replacing
$V_\beta$ in Eq. (\ref{Vbeta1}). The optimization of $V_\beta^{\rm(trial)}$  is performed by requiring that $\partial V_\beta^{\rm(trial)}/\partial M=0$ at $M=M^{\rm(opt)}$ while maintaining a fixed temperature. Reinserting $r$ into the obtained result and setting $\eta=1$ yields
\begin{eqnarray}
M^{\rm(opt)}=\mu-g\int\frac{d^3\vec{k}}{(2\pi)^3}\frac{\varepsilon_k}{\sqrt{\varepsilon_k^2+2M^{\rm(opt)}\varepsilon_k}}\left[\frac{1}{2}\frac{1}{e^{\beta\sqrt{\varepsilon_k^2+2M^{\rm(opt)}\varepsilon_k}}-1}\right]
\end{eqnarray}
 This condition determines $M=M^{\rm(opt)}=g\rho$. As a result, the self-consistent pressure is obtained, maintaining the same form as Eq. (\ref{pressure}), with the condensed density
$\rho_0$ substituted by the total particle density $\rho$. For simplicity the self-consistent pressure is still denoted by ${\cal P}$
\begin{eqnarray}
{\cal P}&=&\frac{\left[M^{\rm(opt)}\right]^2}{2g}-\frac{1}{2}\int\frac{d^3\vec{k}}{(2\pi)^3}\sqrt{\varepsilon_k^2+2g\rho\varepsilon_k}-\frac{1}{\beta}\int\frac{d^3\vec{k}}{(2\pi)^3}\ln\left[1-e^{-\beta\sqrt{\varepsilon_k^2+2g\rho\varepsilon_k}}\right]\nonumber\\
&\equiv &\frac{\left[M^{\rm(opt)}\right]^2}{2g}+{\cal P}^{(0)}_{\rm g}+{\cal P}^{(T)}_{ g}.\label{selfpressure}
\end{eqnarray}
The subsequent steps involve optimizing the CJT effective potential with respect to the condensed density and evaluating the first derivative of the self-consistent pressure with respect to the chemical potential. By combining these results, one obtains
\begin{eqnarray}
\rho-\rho_0=\int\frac{d^3\vec{k}}{(2\pi)^3}\frac{\varepsilon_k+g\rho}{\sqrt{\varepsilon_k^2+2g\rho\varepsilon_k}}\left[\frac{1}{2}+\frac{1}{e^{\beta\sqrt{\varepsilon_k^2+2g\rho\varepsilon_k}}-1}\right].\label{selfrho}
\end{eqnarray}
Equation (\ref{selfrho}) serves as a gap equation, governing the evolution of the system under varying parameters.

\subsection{Transition temperature}

We now calculate the transition temperature of homogeneous repulsive weakly interacting Bose gas. To this end, we first recognize that the first integral in right-hand side of Eq. (\ref{selfrho}), which does not depend explicitly on temperature. Setting $p^2=\hbar^2k^2/2m$, this term can be written in form
\begin{eqnarray}
\frac{1}{2}\int\frac{d^3\vec{k}}{(2\pi)^3}\frac{\varepsilon_k+g\rho}{\sqrt{\varepsilon_k^2+2g\rho\varepsilon_k}}=&&\frac{(2m)^{3/2}}{2\hbar^3}\int\frac{d^3\vec{p}}{(2\pi)^3}\frac{p^2}{\sqrt{p^2(p^2+2g\rho)}}\nonumber\\
&&+\frac{(2m)^{3/2}g\rho}{2\hbar^3}\int\frac{d^3\vec{p}}{(2\pi)^3}\frac{1}{\sqrt{p^2(p^2+2g\rho)}}.\label{integral1}
\end{eqnarray}
It is obvious that the integrals in right-hand side of (\ref{integral1}) are ultraviolet divergent. This divergence is avoidable by means of the dimensional regularization \cite{Andersen2004} and therefore the integrals can be computed. The integral $I_{m,n}$ is
\begin{eqnarray}
        I_{m,n}({\cal M})&=&\int\frac{d^d\kappa}{(2\pi)^d}\frac{\kappa^{2m-n}}{(\kappa^2+{\cal M}^2)^{n/2}}\nonumber\\
        &=&\frac{\Omega_d}{(2\pi)^d}\Lambda^{2\epsilon}{\cal M}^{d+2(m-n)}\frac{\Gamma\left(\frac{d-n}{2}+m\right)\Gamma\left(n-m-\frac{d}{2}\right)}{2\Gamma\left(\frac{n}{2}\right)},\label{eq:tp}
\end{eqnarray}
where $\Gamma(x)$ is the gamma function, $\Omega_d=2\pi^{d/2}/\Gamma(d/2)$ is the surface area of a $d-$dimensional sphere and $\Lambda$ is a renormalization scale that ensures the integral has the correct canonical dimension. Employing (\ref{eq:tp}) the integral (\ref{integral1}) is calculated
\begin{eqnarray}
\frac{1}{2}\int\frac{d^3\vec{k}}{(2\pi)^3}\frac{\varepsilon_k+g\rho}{\sqrt{\varepsilon_k^2+2g\rho\varepsilon_k}}=\frac{(2m)^{3/2}}{24\pi^2\hbar^3}(2g\rho)^{3/2}.\label{result1}
\end{eqnarray}
It is noteworthy that Eq. (\ref{result1}) can be rederived by differentiating Eq. (\ref{zero1}) with respect to the chemical potential. Furthermore, by employing the gas parameter, Eq. (\ref{result1}) can be simplified as follows
\begin{eqnarray}
\frac{1}{2}\int\frac{d^3\vec{k}}{(2\pi)^3}\frac{\varepsilon_k+g\rho}{\sqrt{\varepsilon_k^2+2g\rho\varepsilon_k}}=\frac{8}{3\sqrt{\pi}}\rho\alpha_s^{1/2}.\label{fluctuation}
\end{eqnarray}
It is evident that the right-hand side of Eq. (\ref{fluctuation}) precisely corresponds to the quantum fluctuations of a homogeneous, repulsive, weakly interacting Bose gas. This result was originally derived by Bogoliubov \cite{Bogoliubov} and subsequently confirmed by other researchers \cite{Wu1959,VanThu2022}. Its validity has also been demonstrated through experimental observations \cite{Lopes2017}.

Now we deal with the second integral in the right-hand side of Eq. (\ref{selfrho}). To do this we introduce an auxiliary variable $q^2=\beta\varepsilon_k$. Therefore this integral can be rewritten as follows
\begin{eqnarray}
I&=&\int\frac{d^3\vec{k}}{(2\pi)^3}\frac{\varepsilon_k+g\rho}{\sqrt{\varepsilon_k^2+2g\rho\varepsilon_k}}\frac{1}{e^{\beta\sqrt{\varepsilon_k^2+2g\rho\varepsilon_k}}-1}\nonumber\\
&=&\frac{(2m)^{3/2}}{2\pi^2\hbar^3}\int_0^\infty  dq\frac{q^2(q^2+\beta g\rho)}{\sqrt{q^2(q^2+2\beta g\rho)}}\frac{1}{e^{\sqrt{q^2(q^2+2\beta g\rho)}}-1}.\label{integral2}
\end{eqnarray}
Recall that the system under consideration is the so dilute Bose gas that in region of just below the transition temperature the interaction temperature $T^{\rm(int)}=g\rho/k_B$ is significantly smaller than the transition temperature \cite{Kleinert2005}. Consequently, the parameter $\beta g\rho$ can be regarded as much smaller than unity. To first order in this parameter, the integral (\ref{integral2}) can be approximated as follows
\begin{eqnarray}
I\approx \frac{m^{3/2}}{2\sqrt{2}\pi^{3/2}\hbar^3\beta^{3/2}}\left[\zeta(3/2)+\beta g\rho\zeta(1/2)\right].\label{result2}
\end{eqnarray}
Substituting Eqs. (\ref{fluctuation}) and (\ref{result2}) into Eq. (\ref{selfrho}) yields the expression
\begin{eqnarray}
\frac{\rho_{\rm ex}}{\rho}\equiv\frac{\rho-\rho_0}{\rho}=\frac{8}{3\sqrt{\pi}}\alpha_s^{1/2}+\sqrt{\frac{2m}{\pi\hbar^2}}\frac{\zeta(1/2)a_s}{\beta^{1/2}}+\left(\frac{m}{2\pi}\right)^{3/2}\frac{\zeta(3/2)}{\rho\hbar^3\beta^{3/2}},\label{gap}
\end{eqnarray}
where $\rho_{\rm ex}$ is defined as the non-condensed density. It is notable that the first term on the right-hand side of Eq. (\ref{gap}) corresponds to quantum fluctuations, which reduce the condensed density due to interaction-induced quantum effects.

We now calculate the transition temperature. At the transition point, the condensed density vanishes ($\rho_0=0$). From Eq. (\ref{gap}), the relative shift in the transition temperature, to the lowest order of the $s$-wave scattering length, is given by
\begin{eqnarray}
\frac{\Delta T_C}{T_C^{(0)}}=-\frac{4 \zeta \left(\frac{1}{2}\right)}{3 \zeta(3/2)^{1/3}}\rho^{1/3}a_s.\label{shiftT}
\end{eqnarray}
It is obvious that result (\ref{shiftT}) has the same form as in Eq. (\ref{shift0}) with
\begin{eqnarray}
c=-\frac{4\zeta(1/2)}{3\zeta(3/2)^{1/3}}\approx 1.413\approx\sqrt{2},\label{cnumber}
\end{eqnarray}
and $a=1/3$. These values exhibit a high degree of agreement with our result when compared to those obtained from precise Monte Carlo simulations \cite{Kashurnikov2001,Arnold2001a} $c=1.29\pm 0.05$. Furthermore, it is in excellent concordance with the theoretical result derived using the nonperturbative linear $\delta$-expansion method \cite{SouzaCruz2002} $c=1.32\pm 0.02$. It also aligns closely with other finding in mathematical physics \cite{Napiorkowski2017} $c=1.49$.  Unfortunately, this result indicates a first-order phase transition.

\section{Zero-point energy and thermodynamic properties of the dilute Bose gas\label{sec:3}}
\subsection{Condensed phase}

We now examine the temperature dependence of the zero-point energy and thermodynamic quantities in the condensed phase of the gas. The initial focus is placed on the chemical potential. Inversion of Eq. (\ref{rho}) one has
\begin{eqnarray}
\mu=g\rho+g\int\frac{d^3\vec{k}}{(2\pi)^3}\frac{\varepsilon_k}{\sqrt{\varepsilon_k^2+2\mu\varepsilon_k}}\left[\frac{1}{2}+\frac{1}{e^{\beta\sqrt{\varepsilon_k^2+2\mu\varepsilon_k}}-1}\right].\label{chemical}
\end{eqnarray}
By employing Eqs. (\ref{eq:tp}) and following a procedure analogous to that used for Eq. (\ref{integral2}), the chemical potential can be determined and expressed as a function of temperature to the first order of the scattering length
\begin{eqnarray}
\mu=g\rho\left(1+\frac{32}{3\sqrt{\pi}}\alpha_s^{1/2}\right)+\frac{4\pi\hbar^2\zeta(3/2)a_s}{m\lambda_B^3}.\label{chemical1}
\end{eqnarray}
It is evident that the first term on the right-hand side of Eq. (\ref{chemical1}) represents the contribution from mean-field theory,
$g\rho$, as well as beyond-mean-field effects, which have been previously identified \cite{Haugset1998,VanThu2022}. The temperature dependence is encapsulated in the last two terms, which follow a half-integer power law, in contrast to the integer power law described in the findings of Ref. \cite{Haugset1998}. For the ideal Bose gas, the chemical potential is a function of temperature in form
\begin{eqnarray}
\mu_{\rm{ideal}}=k_BT\ln z,\label{muideal}
\end{eqnarray}
with $z$ being the fugacity, which is unity in the condensed phase \cite{Huang2001}. It is evident that Eq. (\ref{chemical1}) simplifies to Eq. (\ref{muideal}) when $a_s=0$ for the ideal Bose gas.

We now estimate the self-pressure, which is shown in (\ref{selfpressure}). The second term taken with a negative sign, represents the zero-point (vacuum) energy, which generates quantum fluctuations in the ground state \cite{VanThu2022}. In a finite volume, this zero-point energy gives rise to the Casimir effect in BEC(s), a phenomenon that has been extensively investigated (see, for instance, \cite{Thu2020,VanThu2022a})). In the context of the present study, the zero-point energy is given by:
\begin{eqnarray}
V^{(0)}_\beta=-{\cal P}^{(0)}_{\rm g}=\frac{1}{2}\int\frac{d^3\vec{k}}{(2\pi)^3}\sqrt{\varepsilon_k^2+2\mu\varepsilon_k},\label{zeropoint}
\end{eqnarray}
which does not explicitly depend on temperature and is known to exhibit ultraviolet divergence. To obtain a finite physical value, this divergence must be regularized. Fortunately, several regularization methods are available to address this issue \cite{Salasnich2016}. One prominent approach involves using the integral regularization technique developed by ’t Hooft and Veltman \cite{Hooft1972}. Applying this method to the current case yields a finite expression for the zero-point energy:
\begin{eqnarray}
V^{(0)}_\beta=-{\cal P}^{(0)}_{\rm g}=\frac{8m^{3/2}}{15\pi^2\hbar^3}\mu^{5/2}.\label{zero1}
\end{eqnarray}
This result can also be derived using Eq. (\ref{eq:tp}). An alternative approach to resolving the divergence is the momentum-cutoff regularization method. This technique introduces a momentum cutoff $\Lambda$, where the divergent terms are absorbed through appropriate counterterms. The results obtained using this method are consistent with Eq. (\ref{zero1}), but with the chemical potential $\mu$ replaced by its renormalized form $\mu_r=\mu-\frac{g\Lambda^3}{12\pi^2}$. Further details on this approach can be found in Ref. \cite{Schakel2008}. The result (\ref{zero1}) coincides with the original result of Lee and Yang \cite{Lee1960}.

Similarly to Eq. (\ref{result2}), using condition of the dilute gas and the interacting temperature is much smaller than the transition temperature the last term in right-hand side of Eq. (\ref{selfpressure}) is estimated
\begin{eqnarray}
{\cal P}^{(T)}_{ g}=\frac{m^{3/2}\zeta(5/2)}{2\sqrt{2}\pi^{3/2}\hbar^3\beta^{5/2}}-\frac{m^{3/2}\zeta(3/2)}{2\sqrt{2}\pi^{3/2}\hbar^3\beta^{3/2}}\mu.\label{OmegaT}
\end{eqnarray}
This result is different from the one in Ref. \cite{Haugset1998} [cf. Eq. (107)]. Inserting (\ref{zero1}) and (\ref{OmegaT}) into (\ref{selfpressure}), using definition of the de Broglie wavelength
\begin{eqnarray}
\lambda_B=\sqrt{\frac{2\pi\hbar^2}{mk_BT}},\label{Broglie}
\end{eqnarray}
one can write the self-consistent pressure up to the first order of the scattering length as follows
\begin{eqnarray}
{\cal P}=&&\frac{1}{2}g\rho^2\left(1+\frac{128}{15\sqrt{\pi}}\alpha_s^{1/2}\right)+\frac{4\pi\hbar^2\zeta(3/2)}{m\lambda_B^3}\rho a_s-\frac{6\pi\hbar^2\zeta(3/2)^2a_s}{m\lambda_B^6}\nonumber\\
&&+\frac{2\pi\hbar^2\zeta(5/2)}{m\lambda_B^5}.\label{Press1}
\end{eqnarray}
Let us now examine the physical significance of the terms on the right-hand side of Eq. (\ref{Press1}). The first term consists of two parts enclosed within the parentheses and their physical interpretation is not difficult to discern: the first part corresponds to the bulk pressure $g\rho^2/2$, while the second part arises from the zero-point energy. This latter contribution constitutes a contribution beyond the mean-field theory, attributable to quantum fluctuations. This term has been found before in Ref. \cite{VanThu2022} at zero temperature. Two next terms expresses the contribution of both the interatomic interaction and thermal fluctuations on the pressure. The last term is purely attributable to thermal fluctuations. Applying to the perfect Bose gas, i.e., the $s$-wave scattering length vanishes, the right-hand side of Eq. (\ref{Press1}) reduces to this last term. It reveals that the pressure of the perfect Bose gas is proportional to $(k_BT)^{5/2}$ \cite{Huang1987}.

The next thermodynamic quantity to consider is the energy density. At nonzero temperature, the energy density is defined as a Legendre transform of the free-energy density, as given by \cite{VanThu2022,Huang2001}
\begin{eqnarray}
{\cal E}=\mu \rho-{\cal P}.\label{energy}
\end{eqnarray}
Combining this expression with Eq. (\ref{Press1}), the energy density becomes
\begin{eqnarray}
{\cal E}=&&\frac{1}{2}g\rho^2\left(1+\frac{128}{15\sqrt{\pi}}\alpha_s^{1/2}\right)-\frac{4\pi\hbar^2\zeta(3/2)\rho a_s}{m\lambda_B^3}+\frac{2\pi\hbar^2\zeta(3/2)^2a_s}{m\lambda_B^6}\nonumber\\
&&-\frac{2\pi\hbar^2\zeta(5/2)}{m\lambda_B^5}.\label{Ener}
\end{eqnarray}
Likewise the self-consistent pressure, the first term on the right-hand side of Eq. (\ref{Ener}) represents the mean-field and beyond-mean-field contributions to the energy density. These results are consistent with those previously reported in Refs. \cite{Wu1959,Giorgini1999,Carlen2021,VanThu2022,Lee1957} and have been experimentally confirmed \cite{Navon2011}. The contributions arising from thermal fluctuations are captured in the last three terms of the expression.

\subsection{Normal phase}

In the normal phase, that is above the transition temperature, the condensed fraction is zero, and consequently, all atoms remain in excited states. The dispersion relation, as given by Eq. (\ref{Eone}), simplifies to
\begin{eqnarray}
E(k)=\varepsilon_k-\bar\mu,\label{mut1}
\end{eqnarray}
and the propagator, represented by Eq. (\ref{eq:protree}), reduces to
\begin{equation}
D(k)=\frac{1}{\omega_n^2+E^2(k)}\left(
              \begin{array}{cc}
                \varepsilon_k-\bar\mu & \omega_n \\
                -\omega_n &  \varepsilon_k-\bar\mu\\
              \end{array}
            \right).\label{eq:protree1}
\end{equation}
Herein we denote the chemical potential in the normal phase as $\bar\mu$, which is shifted $2g\rho$ relative to that of the ideal Bose gas. The ultraviolet divergence in the zero-point energy can be eliminated through the introduction of a counterterm, ensuring that it does not contribute to the pressure (as discussed in \cite{Haugset1998,Salasnich2016}). The last term on the right-hand side of Equation (23) can be expressed using the polylogarithmic function, defined as ${\rm Li}_n(x)=\sum_{k=1}^\infty \frac{x^k}{n^n}$. The self-consistent pressure is then given by \cite{Mordini2020}
\begin{eqnarray}
{\cal P}=g\rho^2+\frac{2\pi\hbar^2}{m\lambda_B^5}{\rm Li}_{5/2}\left[e^{\beta(\mu-2g\rho)}\right].\label{pressureabove}
\end{eqnarray}
Using (\ref{pressureabove}) and (\ref{density}) one can find the particle density
\begin{eqnarray}
\rho=\frac{{\rm Li}_{5/2}\left[e^{\beta(\mu-2g\rho)}\right]}{\lambda_B^3}.\label{rhoabove}
\end{eqnarray}
Inverting Eq. (\ref{rhoabove}) yields the chemical potential
\begin{eqnarray}
\mu=2g\rho+k_BT\ln{\rm Li}_{3/2}^{-1}(\rho\lambda_B^3).\label{muabove}
\end{eqnarray}
Combining Eqs. (\ref{chemical1}) and (\ref{muabove}) one has
\begin{eqnarray}
\mu=\left\{
      \begin{array}{ll}
        g\rho\left(1+\frac{32}{3\sqrt{\pi}}\alpha_s^{1/2}\right)+\frac{4\pi\hbar^2\zeta(3/2)a_s}{m\lambda_B^3}, & \hbox{$T\leq T_C$;} \\
        2g\rho+k_BT\ln{\rm Li}_{3/2}^{-1}(\rho\lambda_B^3), & \hbox{$T\geq T_C$.}
      \end{array}
    \right.\label{mutotal}
\end{eqnarray}
\begin{figure}[t]
\centering
\includegraphics[width = 0.8\linewidth]{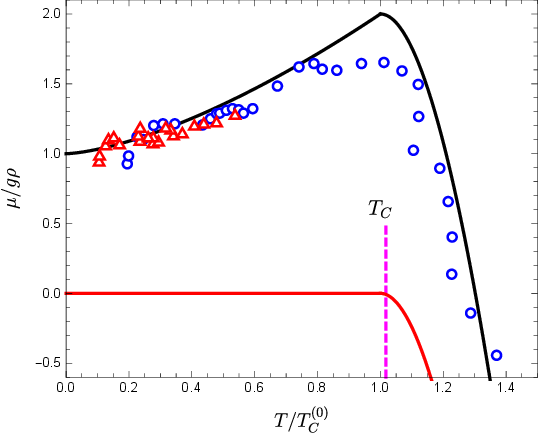}
\caption{(Color online) The evolution of the chemical potential of the sodium 23 gas versus the reduced temperature. The black and red solid curves correspond to chemical potential of the interacting and ideal Bose gas. The blue open dots and red open triangles are the experimental data extracted from Ref. \cite{Mordini2020} for the fitting the
density to the profile calculated at $T=280$ nK and 150 nK, respectively.}\label{fig1}
\end{figure}

The evolution of the chemical potential with respect to the reduced temperature $T/T_C^{(0)}$, as defined in Eq. (\ref{mutotal}), is depicted by the black solid line in Fig. \ref{fig1}. The curve exhibits a continuous variation of the chemical potential across the transition point, further highlighting its non-monotonic behavior. Initially, in the condensed phase, the chemical potential increases monotonically with rising reduced temperature, starting from $g\rho$. It reaches a maximum value of $2g\rho$ (in fact the contribution from the quantum fluctuations is very small) at the critical point, after which it decreases as the temperature continues to increase. The red solid line, for reference, corresponds to the chemical potential of the ideal Bose gas.

We now compare our theoretical results with experimental data. Here we reference the experimental work of Mordini {\it et al.} \cite{Mordini2020}, which investigated a BEC of sodium-23 isotopes confined in an Ioffe-Pritchard trap. In this experiment, parameters were chosen as $m=22.99u, \rho a_s^3=2\times 10^{-6}$. Using the local density approximation, the authors derived certain thermodynamic quantities for the corresponding homogeneous BEC. Experimental data for the chemical potential as a function of reduced temperature is plotted in Fig. \ref{fig1}. The  blue open dots and red open triangles are the experimental data extracted from Ref. \cite{Mordini2020} for the fitting the density to the profile calculated at $T=280$ nK and 150 nK, respectively. These results reveal a strong agreement with our theoretical predictions. An important observation from Fig. \ref{fig1} is that the experimental data confirm the non-monotonic behavior of the chemical potential in the vicinity of the transition point.

Similarly, the energy density above the transition temperature can be calculated. Utilizing Eq. (\ref{energy}), the expression is given by
\begin{eqnarray}
{\cal E}=g\rho^2-\frac{2\pi\hbar^2}{m\lambda_B^5}{\rm Li}_{5/2}\left[e^{\beta(\mu-2g\rho)}\right].\label{enery2}
\end{eqnarray}
This result is consistent with the findings reported in Ref. \cite{Spada2022}.

To end this subsection, we emphasize that the contributions from quantum fluctuations have been ignored in the Eqs. (\ref{pressureabove}), (\ref{muabove}), (\ref{enery2}).

\section{Conclusion and outlook\label{sec:4}}

In the preceding Sections, the fundamental properties of a homogeneous repulsive weakly interacting Bose gas have been analyzed. This investigation utilized the CJT effective action method within the one-loop approximation, combined with the variational perturbation theory in the self-consistent Popov approximation. The principal findings of this study are summarized as follows:

- The relative shift of the transition temperature of the homogeneous repulsive weakly interacting Bose gas with respect to that of the ideal Bose gas has been found in good agreement with that reported by other groups.

- The zero-point energy in the grand canonical ensemble has been examined. It is demonstrated that the ultraviolet divergence can be effectively eliminated using both dimensional regularization and momentum cutoff regularization. These methods yield the same finite result for the zero-point energy, which accounts for the quantum fluctuations superimposed on the ground state of the system.

- The thermodynamic quantities, namely, the chemical potential, pressure and energy density have been investigated in both condensed and non-condensed phases. These results are an improvement of those reported in Refs. \cite{Kleinert2005,Haugset1998}. The comparison of our result for the chemical potential with those reported in experiment shows a excellent agreement.

Notably, despite employing the same self-consistent Popov approximation, our findings differ significantly from those presented in Refs. \cite{Kleinert2005,Haugset1998}. This discrepancy highlights the potential of this approach for further exploring the thermodynamic properties of the interacting Bose gas.

\begin{acknowledgements}
This research is funded by Vietnam National Foundation for Science and Technology Development (NAFOSTED) under grant number 103.01-2023.12.
\end{acknowledgements}

\section*{Conflict of interest}
All of the authors declare that we have no conflict of interest.

\bibliography{popov.bib}

\begin{thebibliography}{52}%
\makeatletter
\providecommand \@ifxundefined [1]{%
 \@ifx{#1\undefined}
}%
\providecommand \@ifnum [1]{%
 \ifnum #1\expandafter \@firstoftwo
 \else \expandafter \@secondoftwo
 \fi
}%
\providecommand \@ifx [1]{%
 \ifx #1\expandafter \@firstoftwo
 \else \expandafter \@secondoftwo
 \fi
}%
\providecommand \natexlab [1]{#1}%
\providecommand \enquote  [1]{``#1''}%
\providecommand \bibnamefont  [1]{#1}%
\providecommand \bibfnamefont [1]{#1}%
\providecommand \citenamefont [1]{#1}%
\providecommand \href@noop [0]{\@secondoftwo}%
\providecommand \href [0]{\begingroup \@sanitize@url \@href}%
\providecommand \@href[1]{\@@startlink{#1}\@@href}%
\providecommand \@@href[1]{\endgroup#1\@@endlink}%
\providecommand \@sanitize@url [0]{\catcode `\\12\catcode `\$12\catcode
  `\&12\catcode `\#12\catcode `\^12\catcode `\_12\catcode `\%12\relax}%
\providecommand \@@startlink[1]{}%
\providecommand \@@endlink[0]{}%
\providecommand \url  [0]{\begingroup\@sanitize@url \@url }%
\providecommand \@url [1]{\endgroup\@href {#1}{\urlprefix }}%
\providecommand \urlprefix  [0]{URL }%
\providecommand \Eprint [0]{\href }%
\providecommand \doibase [0]{https://doi.org/}%
\providecommand \selectlanguage [0]{\@gobble}%
\providecommand \bibinfo  [0]{\@secondoftwo}%
\providecommand \bibfield  [0]{\@secondoftwo}%
\providecommand \translation [1]{[#1]}%
\providecommand \BibitemOpen [0]{}%
\providecommand \bibitemStop [0]{}%
\providecommand \bibitemNoStop [0]{.\EOS\space}%
\providecommand \EOS [0]{\spacefactor3000\relax}%
\providecommand \BibitemShut  [1]{\csname bibitem#1\endcsname}%
\let\auto@bib@innerbib\@empty
\bibitem [{\citenamefont {Bose}(1924)}]{Bose1924}%
  \BibitemOpen
  \bibfield  {author} {\bibinfo {author} {\bibnamefont {Bose}},\ }\bibfield
  {title} {\bibinfo {title} {Plancks gesetz und lichtquantenhypothese},\ }\href
  {https://doi.org/10.1007/bf01327326} {\bibfield  {journal} {\bibinfo
  {journal} {Zeitschrift fur Physik}\ }\textbf {\bibinfo {volume} {26}},\
  \bibinfo {pages} {178} (\bibinfo {year} {1924})}\BibitemShut {NoStop}%
\bibitem [{\citenamefont {Einstein}(1924)}]{Einstein1924}%
  \BibitemOpen
  \bibfield  {author} {\bibinfo {author} {\bibfnamefont {A.}~\bibnamefont
  {Einstein}},\ }\bibfield  {title} {\bibinfo {title} {Quantentheorie des
  einatomigen idealen gases},\ }\href@noop {} {\bibfield  {journal} {\bibinfo
  {journal} {Sitz. Ber. Preuss. Akad. Wiss.}\ }\textbf {\bibinfo {volume}
  {22}},\ \bibinfo {pages} {261} (\bibinfo {year} {1924})}\BibitemShut
  {NoStop}%
\bibitem [{\citenamefont {Anderson}\ \emph {et~al.}(1995)\citenamefont
  {Anderson}, \citenamefont {Ensher}, \citenamefont {Matthews}, \citenamefont
  {Wieman},\ and\ \citenamefont {Cornell}}]{Anderson1995}%
  \BibitemOpen
  \bibfield  {author} {\bibinfo {author} {\bibfnamefont {M.~H.}\ \bibnamefont
  {Anderson}}, \bibinfo {author} {\bibfnamefont {J.~R.}\ \bibnamefont
  {Ensher}}, \bibinfo {author} {\bibfnamefont {M.~R.}\ \bibnamefont
  {Matthews}}, \bibinfo {author} {\bibfnamefont {C.~E.}\ \bibnamefont
  {Wieman}},\ and\ \bibinfo {author} {\bibfnamefont {E.~A.}\ \bibnamefont
  {Cornell}},\ }\bibfield  {title} {\bibinfo {title} {Observation of
  bose-einstein condensation in a dilute atomic vapor},\ }\href
  {https://doi.org/10.1126/science.269.5221.198} {\bibfield  {journal}
  {\bibinfo  {journal} {Science}\ }\textbf {\bibinfo {volume} {269}},\ \bibinfo
  {pages} {198} (\bibinfo {year} {1995})}\BibitemShut {NoStop}%
\bibitem [{\citenamefont {Huang}(1987)}]{Huang1987}%
  \BibitemOpen
  \bibfield  {author} {\bibinfo {author} {\bibfnamefont {K.}~\bibnamefont
  {Huang}},\ }\href@noop {} {\emph {\bibinfo {title} {Statistical
  mechanics}}},\ \bibinfo {edition} {2nd}\ ed.\ (\bibinfo  {publisher}
  {Wiley},\ \bibinfo {address} {New York, NY [u.a.]},\ \bibinfo {year}
  {1987})\BibitemShut {NoStop}%
\bibitem [{\citenamefont {Pethick}(2008)}]{Pethick2008}%
  \BibitemOpen
  \bibfield  {author} {\bibinfo {author} {\bibfnamefont {C.}~\bibnamefont
  {Pethick}},\ }\href@noop {} {\emph {\bibinfo {title} {Bose-Einstein
  condensation in dilute gases}}},\ \bibinfo {edition} {2nd}\ ed.,\ edited by\
  \bibinfo {editor} {\bibfnamefont {H.}~\bibnamefont {Smith}}\ (\bibinfo
  {publisher} {Cambridge University Press},\ \bibinfo {address} {Cambridge ;},\
  \bibinfo {year} {2008})\ \bibinfo {note} {includes bibliographical references
  and index.}\BibitemShut {Stop}%
\bibitem [{\citenamefont {Wu}(1959)}]{Wu1959}%
  \BibitemOpen
  \bibfield  {author} {\bibinfo {author} {\bibfnamefont {T.~T.}\ \bibnamefont
  {Wu}},\ }\bibfield  {title} {\bibinfo {title} {Ground state of a bose system
  of hard spheres},\ }\href {https://doi.org/10.1103/physrev.115.1390}
  {\bibfield  {journal} {\bibinfo  {journal} {Physical Review}\ }\textbf
  {\bibinfo {volume} {115}},\ \bibinfo {pages} {1390} (\bibinfo {year}
  {1959})}\BibitemShut {NoStop}%
\bibitem [{\citenamefont {Giorgini}\ \emph {et~al.}(1999)\citenamefont
  {Giorgini}, \citenamefont {Boronat},\ and\ \citenamefont
  {Casulleras}}]{Giorgini1999}%
  \BibitemOpen
  \bibfield  {author} {\bibinfo {author} {\bibfnamefont {S.}~\bibnamefont
  {Giorgini}}, \bibinfo {author} {\bibfnamefont {J.}~\bibnamefont {Boronat}},\
  and\ \bibinfo {author} {\bibfnamefont {J.}~\bibnamefont {Casulleras}},\
  }\bibfield  {title} {\bibinfo {title} {Ground state of a homogeneous bose
  gas: A diffusion monte carlo calculation},\ }\href
  {https://doi.org/10.1103/physreva.60.5129} {\bibfield  {journal} {\bibinfo
  {journal} {Physical Review A}\ }\textbf {\bibinfo {volume} {60}},\ \bibinfo
  {pages} {5129} (\bibinfo {year} {1999})}\BibitemShut {NoStop}%
\bibitem [{\citenamefont {Carlen}\ \emph {et~al.}(2021)\citenamefont {Carlen},
  \citenamefont {Holzmann}, \citenamefont {Jauslin},\ and\ \citenamefont
  {Lieb}}]{Carlen2021}%
  \BibitemOpen
  \bibfield  {author} {\bibinfo {author} {\bibfnamefont {E.~A.}\ \bibnamefont
  {Carlen}}, \bibinfo {author} {\bibfnamefont {M.}~\bibnamefont {Holzmann}},
  \bibinfo {author} {\bibfnamefont {I.}~\bibnamefont {Jauslin}},\ and\ \bibinfo
  {author} {\bibfnamefont {E.~H.}\ \bibnamefont {Lieb}},\ }\bibfield  {title}
  {\bibinfo {title} {Simplified approach to the repulsive bose gas from low to
  high densities and its numerical accuracy},\ }\href
  {https://doi.org/10.1103/physreva.103.053309} {\bibfield  {journal} {\bibinfo
   {journal} {Physical Review A}\ }\textbf {\bibinfo {volume} {103}},\ \bibinfo
  {pages} {053309} (\bibinfo {year} {2021})}\BibitemShut {NoStop}%
\bibitem [{\citenamefont {Toyoda}(1982)}]{Toyoda1982}%
  \BibitemOpen
  \bibfield  {author} {\bibinfo {author} {\bibfnamefont {T.}~\bibnamefont
  {Toyoda}},\ }\bibfield  {title} {\bibinfo {title} {A microscopic theory of
  the lambda transition},\ }\href
  {https://doi.org/10.1016/0003-4916(82)90277-9} {\bibfield  {journal}
  {\bibinfo  {journal} {Annals of Physics}\ }\textbf {\bibinfo {volume}
  {141}},\ \bibinfo {pages} {154} (\bibinfo {year} {1982})}\BibitemShut
  {NoStop}%
\bibitem [{\citenamefont {Grüter}\ \emph {et~al.}(1997)\citenamefont
  {Grüter}, \citenamefont {Ceperley},\ and\ \citenamefont
  {Laloë}}]{Grueter1997}%
  \BibitemOpen
  \bibfield  {author} {\bibinfo {author} {\bibfnamefont {P.}~\bibnamefont
  {Grüter}}, \bibinfo {author} {\bibfnamefont {D.}~\bibnamefont {Ceperley}},\
  and\ \bibinfo {author} {\bibfnamefont {F.}~\bibnamefont {Laloë}},\
  }\bibfield  {title} {\bibinfo {title} {Critical temperature of bose-einstein
  condensation of hard-sphere gases},\ }\href
  {https://doi.org/10.1103/physrevlett.79.3549} {\bibfield  {journal} {\bibinfo
   {journal} {Physical Review Letters}\ }\textbf {\bibinfo {volume} {79}},\
  \bibinfo {pages} {3549} (\bibinfo {year} {1997})}\BibitemShut {NoStop}%
\bibitem [{\citenamefont {Arnold}\ and\ \citenamefont
  {Moore}(2001{\natexlab{a}})}]{Arnold2001}%
  \BibitemOpen
  \bibfield  {author} {\bibinfo {author} {\bibfnamefont {P.}~\bibnamefont
  {Arnold}}\ and\ \bibinfo {author} {\bibfnamefont {G.~D.}\ \bibnamefont
  {Moore}},\ }\bibfield  {title} {\bibinfo {title} {Monte carlo simulation of
  $o(2)\varphi^4$ field theory in three dimensions},\ }\href
  {https://doi.org/10.1103/physreve.64.066113} {\bibfield  {journal} {\bibinfo
  {journal} {Physical Review E}\ }\textbf {\bibinfo {volume} {64}},\ \bibinfo
  {pages} {066113} (\bibinfo {year} {2001}{\natexlab{a}})}\BibitemShut
  {NoStop}%
\bibitem [{\citenamefont {Napiórkowski}\ \emph {et~al.}(2017)\citenamefont
  {Napiórkowski}, \citenamefont {Reuvers},\ and\ \citenamefont
  {Solovej}}]{Napiorkowski2017}%
  \BibitemOpen
  \bibfield  {author} {\bibinfo {author} {\bibfnamefont {M.}~\bibnamefont
  {Napiórkowski}}, \bibinfo {author} {\bibfnamefont {R.}~\bibnamefont
  {Reuvers}},\ and\ \bibinfo {author} {\bibfnamefont {J.~P.}\ \bibnamefont
  {Solovej}},\ }\bibfield  {title} {\bibinfo {title} {The bogoliubov free
  energy functional ii: The dilute limit},\ }\href
  {https://doi.org/10.1007/s00220-017-3064-x} {\bibfield  {journal} {\bibinfo
  {journal} {Communications in Mathematical Physics}\ }\textbf {\bibinfo
  {volume} {360}},\ \bibinfo {pages} {347} (\bibinfo {year}
  {2017})}\BibitemShut {NoStop}%
\bibitem [{\citenamefont {Kleinert}\ \emph {et~al.}(2005)\citenamefont
  {Kleinert}, \citenamefont {Schmidt},\ and\ \citenamefont
  {Pelster}}]{Kleinert2005}%
  \BibitemOpen
  \bibfield  {author} {\bibinfo {author} {\bibfnamefont {H.}~\bibnamefont
  {Kleinert}}, \bibinfo {author} {\bibfnamefont {S.}~\bibnamefont {Schmidt}},\
  and\ \bibinfo {author} {\bibfnamefont {A.}~\bibnamefont {Pelster}},\
  }\bibfield  {title} {\bibinfo {title} {Quantum phase diagram for homogeneous
  bose‐einstein condensate},\ }\href
  {https://doi.org/10.1002/andp.20055170402} {\bibfield  {journal} {\bibinfo
  {journal} {Annalen der Physik}\ }\textbf {\bibinfo {volume} {517}},\ \bibinfo
  {pages} {214} (\bibinfo {year} {2005})}\BibitemShut {NoStop}%
\bibitem [{\citenamefont {Huang}(1999)}]{Huang1999}%
  \BibitemOpen
  \bibfield  {author} {\bibinfo {author} {\bibfnamefont {K.}~\bibnamefont
  {Huang}},\ }\bibfield  {title} {\bibinfo {title} {Transition temperature of a
  uniform imperfect bose gas},\ }\href
  {https://doi.org/10.1103/physrevlett.83.3770} {\bibfield  {journal} {\bibinfo
   {journal} {Physical Review Letters}\ }\textbf {\bibinfo {volume} {83}},\
  \bibinfo {pages} {3770} (\bibinfo {year} {1999})}\BibitemShut {NoStop}%
\bibitem [{\citenamefont {Seiringer}\ and\ \citenamefont
  {Ueltschi}(2009)}]{Seiringer2009}%
  \BibitemOpen
  \bibfield  {author} {\bibinfo {author} {\bibfnamefont {R.}~\bibnamefont
  {Seiringer}}\ and\ \bibinfo {author} {\bibfnamefont {D.}~\bibnamefont
  {Ueltschi}},\ }\bibfield  {title} {\bibinfo {title} {Rigorous upper bound on
  the critical temperature of dilute bose gases},\ }\href
  {https://doi.org/10.1103/physrevb.80.014502} {\bibfield  {journal} {\bibinfo
  {journal} {Physical Review B}\ }\textbf {\bibinfo {volume} {80}},\ \bibinfo
  {pages} {014502} (\bibinfo {year} {2009})}\BibitemShut {NoStop}%
\bibitem [{\citenamefont {Wilkens}\ \emph {et~al.}(2000)\citenamefont
  {Wilkens}, \citenamefont {Illuminati},\ and\ \citenamefont
  {Krämer}}]{Wilkens2000}%
  \BibitemOpen
  \bibfield  {author} {\bibinfo {author} {\bibfnamefont {M.}~\bibnamefont
  {Wilkens}}, \bibinfo {author} {\bibfnamefont {F.}~\bibnamefont
  {Illuminati}},\ and\ \bibinfo {author} {\bibfnamefont {M.}~\bibnamefont
  {Krämer}},\ }\bibfield  {title} {\bibinfo {title} {Transition temperature of
  the weakly interacting bose gas: perturbative solution of the crossover
  equations in the canonical ensemble},\ }\href
  {https://doi.org/10.1088/0953-4075/33/20/10j} {\bibfield  {journal} {\bibinfo
   {journal} {Journal of Physics B: Atomic, Molecular and Optical Physics}\
  }\textbf {\bibinfo {volume} {33}},\ \bibinfo {pages} {L779} (\bibinfo {year}
  {2000})}\BibitemShut {NoStop}%
\bibitem [{\citenamefont {Davis}\ and\ \citenamefont
  {Morgan}(2003)}]{Davis2003}%
  \BibitemOpen
  \bibfield  {author} {\bibinfo {author} {\bibfnamefont {M.~J.}\ \bibnamefont
  {Davis}}\ and\ \bibinfo {author} {\bibfnamefont {S.~A.}\ \bibnamefont
  {Morgan}},\ }\bibfield  {title} {\bibinfo {title} {Microcanonical temperature
  for a classical field: Application to bose-einstein condensation},\ }\href
  {https://doi.org/10.1103/physreva.68.053615} {\bibfield  {journal} {\bibinfo
  {journal} {Physical Review A}\ }\textbf {\bibinfo {volume} {68}},\ \bibinfo
  {pages} {053615} (\bibinfo {year} {2003})}\BibitemShut {NoStop}%
\bibitem [{\citenamefont {de~Souza~Cruz}\ \emph {et~al.}(2001)\citenamefont
  {de~Souza~Cruz}, \citenamefont {Pinto},\ and\ \citenamefont
  {Ramos}}]{SouzaCruz2001}%
  \BibitemOpen
  \bibfield  {author} {\bibinfo {author} {\bibfnamefont {F.~F.}\ \bibnamefont
  {de~Souza~Cruz}}, \bibinfo {author} {\bibfnamefont {M.~B.}\ \bibnamefont
  {Pinto}},\ and\ \bibinfo {author} {\bibfnamefont {R.~O.}\ \bibnamefont
  {Ramos}},\ }\bibfield  {title} {\bibinfo {title} {Transition temperature for
  weakly interacting homogeneous bose gases},\ }\href
  {https://doi.org/10.1103/physrevb.64.014515} {\bibfield  {journal} {\bibinfo
  {journal} {Physical Review B}\ }\textbf {\bibinfo {volume} {64}},\ \bibinfo
  {pages} {014515} (\bibinfo {year} {2001})}\BibitemShut {NoStop}%
\bibitem [{\citenamefont {Stoof}(1992)}]{Stoof1992}%
  \BibitemOpen
  \bibfield  {author} {\bibinfo {author} {\bibfnamefont {H.~T.~C.}\
  \bibnamefont {Stoof}},\ }\bibfield  {title} {\bibinfo {title} {Nucleation of
  bose-einstein condensation},\ }\href
  {https://doi.org/10.1103/physreva.45.8398} {\bibfield  {journal} {\bibinfo
  {journal} {Physical Review A}\ }\textbf {\bibinfo {volume} {45}},\ \bibinfo
  {pages} {8398} (\bibinfo {year} {1992})}\BibitemShut {NoStop}%
\bibitem [{\citenamefont {Reppy}\ \emph {et~al.}(2000)\citenamefont {Reppy},
  \citenamefont {Crooker}, \citenamefont {Hebral}, \citenamefont {Corwin},
  \citenamefont {He},\ and\ \citenamefont {Zassenhaus}}]{Reppy2000}%
  \BibitemOpen
  \bibfield  {author} {\bibinfo {author} {\bibfnamefont {J.~D.}\ \bibnamefont
  {Reppy}}, \bibinfo {author} {\bibfnamefont {B.~C.}\ \bibnamefont {Crooker}},
  \bibinfo {author} {\bibfnamefont {B.}~\bibnamefont {Hebral}}, \bibinfo
  {author} {\bibfnamefont {A.~D.}\ \bibnamefont {Corwin}}, \bibinfo {author}
  {\bibfnamefont {J.}~\bibnamefont {He}},\ and\ \bibinfo {author}
  {\bibfnamefont {G.~M.}\ \bibnamefont {Zassenhaus}},\ }\bibfield  {title}
  {\bibinfo {title} {Density dependence of the transition temperature in a
  homogeneous bose-einstein condensate},\ }\href
  {https://doi.org/10.1103/physrevlett.84.2060} {\bibfield  {journal} {\bibinfo
   {journal} {Physical Review Letters}\ }\textbf {\bibinfo {volume} {84}},\
  \bibinfo {pages} {2060} (\bibinfo {year} {2000})}\BibitemShut {NoStop}%
\bibitem [{\citenamefont {Al-Sugheir}\ \emph {et~al.}(2023)\citenamefont
  {Al-Sugheir}, \citenamefont {Esbaih}, \citenamefont {Joudeh},\ and\
  \citenamefont {Ghassib}}]{AlSugheir2023}%
  \BibitemOpen
  \bibfield  {author} {\bibinfo {author} {\bibfnamefont {M.~K.}\ \bibnamefont
  {Al-Sugheir}}, \bibinfo {author} {\bibfnamefont {D.~E.}\ \bibnamefont
  {Esbaih}}, \bibinfo {author} {\bibfnamefont {B.~R.}\ \bibnamefont {Joudeh}},\
  and\ \bibinfo {author} {\bibfnamefont {H.~B.}\ \bibnamefont {Ghassib}},\
  }\bibfield  {title} {\bibinfo {title} {Thermodynamic properties and effective
  mass of a weakly-interacting bose gas using the static fluctuation
  approximation},\ }\href {https://doi.org/10.1016/j.physb.2023.414943}
  {\bibfield  {journal} {\bibinfo  {journal} {Physica B: Condensed Matter}\
  }\textbf {\bibinfo {volume} {661}},\ \bibinfo {pages} {414943} (\bibinfo
  {year} {2023})}\BibitemShut {NoStop}%
\bibitem [{\citenamefont {Vianello}\ and\ \citenamefont
  {Salasnich}(2024)}]{Vianello2024}%
  \BibitemOpen
  \bibfield  {author} {\bibinfo {author} {\bibfnamefont {C.}~\bibnamefont
  {Vianello}}\ and\ \bibinfo {author} {\bibfnamefont {L.}~\bibnamefont
  {Salasnich}},\ }\bibfield  {title} {\bibinfo {title} {Condensate and
  superfluid fraction of homogeneous bose gases in a self-consistent popov
  approximation},\ }\bibfield  {journal} {\bibinfo  {journal} {Scientific
  Reports}\ }\textbf {\bibinfo {volume} {14}},\ \href
  {https://doi.org/10.1038/s41598-024-65897-2} {10.1038/s41598-024-65897-2}
  (\bibinfo {year} {2024})\BibitemShut {NoStop}%
\bibitem [{\citenamefont {Kashurnikov}\ \emph {et~al.}(2001)\citenamefont
  {Kashurnikov}, \citenamefont {Prokof’ev},\ and\ \citenamefont
  {Svistunov}}]{Kashurnikov2001}%
  \BibitemOpen
  \bibfield  {author} {\bibinfo {author} {\bibfnamefont {V.~A.}\ \bibnamefont
  {Kashurnikov}}, \bibinfo {author} {\bibfnamefont {N.~V.}\ \bibnamefont
  {Prokof’ev}},\ and\ \bibinfo {author} {\bibfnamefont {B.~V.}\ \bibnamefont
  {Svistunov}},\ }\bibfield  {title} {\bibinfo {title} {Critical temperature
  shift in weakly interacting bose gas},\ }\href
  {https://doi.org/10.1103/physrevlett.87.120402} {\bibfield  {journal}
  {\bibinfo  {journal} {Physical Review Letters}\ }\textbf {\bibinfo {volume}
  {87}},\ \bibinfo {pages} {120402} (\bibinfo {year} {2001})}\BibitemShut
  {NoStop}%
\bibitem [{\citenamefont {Arnold}\ and\ \citenamefont
  {Moore}(2001{\natexlab{b}})}]{Arnold2001a}%
  \BibitemOpen
  \bibfield  {author} {\bibinfo {author} {\bibfnamefont {P.}~\bibnamefont
  {Arnold}}\ and\ \bibinfo {author} {\bibfnamefont {G.}~\bibnamefont {Moore}},\
  }\bibfield  {title} {\bibinfo {title} {Bec transition temperature of a dilute
  homogeneous imperfect bose gas},\ }\href
  {https://doi.org/10.1103/physrevlett.87.120401} {\bibfield  {journal}
  {\bibinfo  {journal} {Physical Review Letters}\ }\textbf {\bibinfo {volume}
  {87}},\ \bibinfo {pages} {120401} (\bibinfo {year}
  {2001}{\natexlab{b}})}\BibitemShut {NoStop}%
\bibitem [{\citenamefont {Griffin}(2009)}]{Griffin2009}%
  \BibitemOpen
  \bibfield  {author} {\bibinfo {author} {\bibfnamefont {A.}~\bibnamefont
  {Griffin}},\ }\href@noop {} {\emph {\bibinfo {title} {Bose-condensed gases at
  finite temperatures}}},\ edited by\ \bibinfo {editor} {\bibfnamefont
  {E.}~\bibnamefont {Zaremba}}\ and\ \bibinfo {editor} {\bibfnamefont
  {T.}~\bibnamefont {Nikuni}}\ (\bibinfo  {publisher} {Cambridge University
  Press},\ \bibinfo {address} {Cambridge},\ \bibinfo {year} {2009})\ \bibinfo
  {note} {includes bibliographical references (pages 451-458) and
  index}\BibitemShut {NoStop}%
\bibitem [{\citenamefont {Griffin}(1996)}]{Griffin1996}%
  \BibitemOpen
  \bibfield  {author} {\bibinfo {author} {\bibfnamefont {A.}~\bibnamefont
  {Griffin}},\ }\bibfield  {title} {\bibinfo {title} {Conserving and gapless
  approximations for an inhomogeneous bose gas at finite temperatures},\ }\href
  {https://doi.org/10.1103/physrevb.53.9341} {\bibfield  {journal} {\bibinfo
  {journal} {Physical Review B}\ }\textbf {\bibinfo {volume} {53}},\ \bibinfo
  {pages} {9341} (\bibinfo {year} {1996})}\BibitemShut {NoStop}%
\bibitem [{\citenamefont {Andersen}(2004)}]{Andersen2004}%
  \BibitemOpen
  \bibfield  {author} {\bibinfo {author} {\bibfnamefont {J.}~\bibnamefont
  {Andersen}},\ }\bibfield  {title} {\bibinfo {title} {Theory of the weakly
  interacting bose gas},\ }\href {https://doi.org/10.1103/revmodphys.76.599}
  {\bibfield  {journal} {\bibinfo  {journal} {Reviews of Modern Physics}\
  }\textbf {\bibinfo {volume} {76}},\ \bibinfo {pages} {599} (\bibinfo {year}
  {2004})}\BibitemShut {NoStop}%
\bibitem [{\citenamefont {Shi}(1998)}]{Shi1998}%
  \BibitemOpen
  \bibfield  {author} {\bibinfo {author} {\bibfnamefont {H.}~\bibnamefont
  {Shi}},\ }\bibfield  {title} {\bibinfo {title} {Finite-temperature
  excitations in a dilute bose-condensed gas},\ }\href
  {https://doi.org/10.1016/s0370-1573(98)00015-5} {\bibfield  {journal}
  {\bibinfo  {journal} {Physics Reports}\ }\textbf {\bibinfo {volume} {304}},\
  \bibinfo {pages} {1} (\bibinfo {year} {1998})}\BibitemShut {NoStop}%
\bibitem [{\citenamefont {Yukalov}\ and\ \citenamefont
  {Kleinert}(2006)}]{Yukalov2006}%
  \BibitemOpen
  \bibfield  {author} {\bibinfo {author} {\bibfnamefont {V.~I.}\ \bibnamefont
  {Yukalov}}\ and\ \bibinfo {author} {\bibfnamefont {H.}~\bibnamefont
  {Kleinert}},\ }\bibfield  {title} {\bibinfo {title} {Gapless
  hartree-fock-bogoliubov approximation for bose gases},\ }\href
  {https://doi.org/10.1103/physreva.73.063612} {\bibfield  {journal} {\bibinfo
  {journal} {Physical Review A}\ }\textbf {\bibinfo {volume} {73}},\ \bibinfo
  {pages} {063612} (\bibinfo {year} {2006})}\BibitemShut {NoStop}%
\bibitem [{\citenamefont {Haugset}\ \emph {et~al.}(1998)\citenamefont
  {Haugset}, \citenamefont {Haugerud},\ and\ \citenamefont
  {Ravndal}}]{Haugset1998}%
  \BibitemOpen
  \bibfield  {author} {\bibinfo {author} {\bibfnamefont {T.}~\bibnamefont
  {Haugset}}, \bibinfo {author} {\bibfnamefont {H.}~\bibnamefont {Haugerud}},\
  and\ \bibinfo {author} {\bibfnamefont {F.}~\bibnamefont {Ravndal}},\
  }\bibfield  {title} {\bibinfo {title} {Thermodynamics of a weakly interacting
  bose–einstein gas},\ }\href {https://doi.org/10.1006/aphy.1998.5795}
  {\bibfield  {journal} {\bibinfo  {journal} {Annals of Physics}\ }\textbf
  {\bibinfo {volume} {266}},\ \bibinfo {pages} {27} (\bibinfo {year}
  {1998})}\BibitemShut {NoStop}%
\bibitem [{\citenamefont {Goldstone}(1961)}]{Goldstone1961}%
  \BibitemOpen
  \bibfield  {author} {\bibinfo {author} {\bibfnamefont {J.}~\bibnamefont
  {Goldstone}},\ }\bibfield  {title} {\bibinfo {title} {Field theories with
  superconductor solutions},\ }\href {https://doi.org/10.1007/bf02812722}
  {\bibfield  {journal} {\bibinfo  {journal} {Il Nuovo Cimento}\ }\textbf
  {\bibinfo {volume} {19}},\ \bibinfo {pages} {154} (\bibinfo {year}
  {1961})}\BibitemShut {NoStop}%
\bibitem [{\citenamefont {Hugenholtz}\ and\ \citenamefont
  {Pines}(1959)}]{Hugenholtz1959}%
  \BibitemOpen
  \bibfield  {author} {\bibinfo {author} {\bibfnamefont {N.~M.}\ \bibnamefont
  {Hugenholtz}}\ and\ \bibinfo {author} {\bibfnamefont {D.}~\bibnamefont
  {Pines}},\ }\bibfield  {title} {\bibinfo {title} {Ground-state energy and
  excitation spectrum of a system of interacting bosons},\ }\href
  {https://doi.org/10.1103/physrev.116.489} {\bibfield  {journal} {\bibinfo
  {journal} {Physical Review}\ }\textbf {\bibinfo {volume} {116}},\ \bibinfo
  {pages} {489} (\bibinfo {year} {1959})}\BibitemShut {NoStop}%
\bibitem [{\citenamefont {Hohenberg}\ and\ \citenamefont
  {Martin}(1965)}]{Hohenberg1965}%
  \BibitemOpen
  \bibfield  {author} {\bibinfo {author} {\bibfnamefont {P.}~\bibnamefont
  {Hohenberg}}\ and\ \bibinfo {author} {\bibfnamefont {P.}~\bibnamefont
  {Martin}},\ }\bibfield  {title} {\bibinfo {title} {Microscopic theory of
  superfluid helium},\ }\href {https://doi.org/10.1016/0003-4916(65)90280-0}
  {\bibfield  {journal} {\bibinfo  {journal} {Annals of Physics}\ }\textbf
  {\bibinfo {volume} {34}},\ \bibinfo {pages} {291} (\bibinfo {year}
  {1965})}\BibitemShut {NoStop}%
\bibitem [{\citenamefont {Thu}\ and\ \citenamefont {Pham}(2024)}]{Thu2024}%
  \BibitemOpen
  \bibfield  {author} {\bibinfo {author} {\bibfnamefont {N.~V.}\ \bibnamefont
  {Thu}}\ and\ \bibinfo {author} {\bibfnamefont {D.~T.}\ \bibnamefont {Pham}},\
  }\bibfield  {title} {\bibinfo {title} {Effect of nonzero temperature to
  non-condensed fraction of a homogeneous dilute weakly interacting bose gas},\
  }\href {https://doi.org/10.1016/j.physleta.2024.129787} {\bibfield  {journal}
  {\bibinfo  {journal} {Physics Letters A}\ }\textbf {\bibinfo {volume}
  {523}},\ \bibinfo {pages} {129787} (\bibinfo {year} {2024})}\BibitemShut
  {NoStop}%
\bibitem [{\citenamefont {Van~Thu}\ and\ \citenamefont
  {Theu}(2017)}]{VanThu2017}%
  \BibitemOpen
  \bibfield  {author} {\bibinfo {author} {\bibfnamefont {N.}~\bibnamefont
  {Van~Thu}}\ and\ \bibinfo {author} {\bibfnamefont {L.~T.}\ \bibnamefont
  {Theu}},\ }\bibfield  {title} {\bibinfo {title} {Casimir force of
  two-component bose–einstein condensates confined by a parallel plate
  geometry},\ }\href {https://doi.org/10.1007/s10955-017-1800-4} {\bibfield
  {journal} {\bibinfo  {journal} {Journal of Statistical Physics}\ }\textbf
  {\bibinfo {volume} {168}},\ \bibinfo {pages} {1} (\bibinfo {year}
  {2017})}\BibitemShut {NoStop}%
\bibitem [{\citenamefont {Schmitt}(2010)}]{Schmitt2010}%
  \BibitemOpen
  \bibfield  {author} {\bibinfo {author} {\bibfnamefont {A.}~\bibnamefont
  {Schmitt}},\ }\href {https://doi.org/10.1007/978-3-642-12866-0} {\emph
  {\bibinfo {title} {Dense Matter in Compact Stars}}}\ (\bibinfo  {publisher}
  {Springer Berlin Heidelberg},\ \bibinfo {year} {2010})\BibitemShut {NoStop}%
\bibitem [{\citenamefont {Thu}(2023)}]{Thu2023}%
  \BibitemOpen
  \bibfield  {author} {\bibinfo {author} {\bibfnamefont {N.~V.}\ \bibnamefont
  {Thu}},\ }\bibfield  {title} {\bibinfo {title} {Non-condensate fraction of a
  weakly interacting bose gas confined between two parallel plates within
  improved hartree-fock approximation at zero temperature},\ }\href
  {https://doi.org/10.1016/j.physleta.2023.129099} {\bibfield  {journal}
  {\bibinfo  {journal} {Physics Letters A}\ }\textbf {\bibinfo {volume}
  {486}},\ \bibinfo {pages} {129099} (\bibinfo {year} {2023})}\BibitemShut
  {NoStop}%
\bibitem [{\citenamefont {Bogoliubov}(1947)}]{Bogoliubov}%
  \BibitemOpen
  \bibfield  {author} {\bibinfo {author} {\bibfnamefont {N.~N.}\ \bibnamefont
  {Bogoliubov}},\ }\bibfield  {title} {\bibinfo {title} {On the theory of
  superfluidity},\ }\href@noop {} {\bibfield  {journal} {\bibinfo  {journal}
  {Journal of Physics}\ }\textbf {\bibinfo {volume} {11}},\ \bibinfo {pages}
  {23} (\bibinfo {year} {1947})}\BibitemShut {NoStop}%
\bibitem [{\citenamefont {Van~Thu}\ and\ \citenamefont
  {Berx}(2022)}]{VanThu2022}%
  \BibitemOpen
  \bibfield  {author} {\bibinfo {author} {\bibfnamefont {N.}~\bibnamefont
  {Van~Thu}}\ and\ \bibinfo {author} {\bibfnamefont {J.}~\bibnamefont {Berx}},\
  }\bibfield  {title} {\bibinfo {title} {The condensed fraction of a
  homogeneous dilute bose gas within the improved hartree–fock
  approximation},\ }\bibfield  {journal} {\bibinfo  {journal} {Journal of
  Statistical Physics}\ }\textbf {\bibinfo {volume} {188}},\ \href
  {https://doi.org/10.1007/s10955-022-02944-0} {10.1007/s10955-022-02944-0}
  (\bibinfo {year} {2022})\BibitemShut {NoStop}%
\bibitem [{\citenamefont {Lopes}\ \emph {et~al.}(2017)\citenamefont {Lopes},
  \citenamefont {Eigen}, \citenamefont {Navon}, \citenamefont {Clément},
  \citenamefont {Smith},\ and\ \citenamefont {Hadzibabic}}]{Lopes2017}%
  \BibitemOpen
  \bibfield  {author} {\bibinfo {author} {\bibfnamefont {R.}~\bibnamefont
  {Lopes}}, \bibinfo {author} {\bibfnamefont {C.}~\bibnamefont {Eigen}},
  \bibinfo {author} {\bibfnamefont {N.}~\bibnamefont {Navon}}, \bibinfo
  {author} {\bibfnamefont {D.}~\bibnamefont {Clément}}, \bibinfo {author}
  {\bibfnamefont {R.~P.}\ \bibnamefont {Smith}},\ and\ \bibinfo {author}
  {\bibfnamefont {Z.}~\bibnamefont {Hadzibabic}},\ }\bibfield  {title}
  {\bibinfo {title} {Quantum depletion of a homogeneous bose-einstein
  condensate},\ }\href {https://doi.org/10.1103/physrevlett.119.190404}
  {\bibfield  {journal} {\bibinfo  {journal} {Physical Review Letters}\
  }\textbf {\bibinfo {volume} {119}},\ \bibinfo {pages} {190404} (\bibinfo
  {year} {2017})}\BibitemShut {NoStop}%
\bibitem [{\citenamefont {de~Souza~Cruz}\ \emph {et~al.}(2002)\citenamefont
  {de~Souza~Cruz}, \citenamefont {Pinto}, \citenamefont {Ramos},\ and\
  \citenamefont {Sena}}]{SouzaCruz2002}%
  \BibitemOpen
  \bibfield  {author} {\bibinfo {author} {\bibfnamefont {F.~F.}\ \bibnamefont
  {de~Souza~Cruz}}, \bibinfo {author} {\bibfnamefont {M.~B.}\ \bibnamefont
  {Pinto}}, \bibinfo {author} {\bibfnamefont {R.~O.}\ \bibnamefont {Ramos}},\
  and\ \bibinfo {author} {\bibfnamefont {P.}~\bibnamefont {Sena}},\ }\bibfield
  {title} {\bibinfo {title} {Higher-order evaluation of the critical
  temperature for interacting homogeneous dilute bose gases},\ }\href
  {https://doi.org/10.1103/physreva.65.053613} {\bibfield  {journal} {\bibinfo
  {journal} {Physical Review A}\ }\textbf {\bibinfo {volume} {65}},\ \bibinfo
  {pages} {053613} (\bibinfo {year} {2002})}\BibitemShut {NoStop}%
\bibitem [{\citenamefont {Huang}(2001)}]{Huang2001}%
  \BibitemOpen
  \bibfield  {author} {\bibinfo {author} {\bibfnamefont {K.}~\bibnamefont
  {Huang}},\ }\href@noop {} {\emph {\bibinfo {title} {Introduction to
  statistical physics}}}\ (\bibinfo  {publisher} {CRC Press},\ \bibinfo
  {address} {Boca Raton, Fla. [u.a.]},\ \bibinfo {year} {2001})\BibitemShut
  {NoStop}%
\bibitem [{\citenamefont {Thu}\ and\ \citenamefont {Song}(2020)}]{Thu2020}%
  \BibitemOpen
  \bibfield  {author} {\bibinfo {author} {\bibfnamefont {N.~V.}\ \bibnamefont
  {Thu}}\ and\ \bibinfo {author} {\bibfnamefont {P.~T.}\ \bibnamefont {Song}},\
  }\bibfield  {title} {\bibinfo {title} {Casimir effect in a weakly interacting
  bose gas confined by a parallel plate geometry in improved hartree–fock
  approximation},\ }\href {https://doi.org/10.1016/j.physa.2019.123018}
  {\bibfield  {journal} {\bibinfo  {journal} {Physica A: Statistical Mechanics
  and its Applications}\ }\textbf {\bibinfo {volume} {540}},\ \bibinfo {pages}
  {123018} (\bibinfo {year} {2020})}\BibitemShut {NoStop}%
\bibitem [{\citenamefont {Van~Thu}(2022)}]{VanThu2022a}%
  \BibitemOpen
  \bibfield  {author} {\bibinfo {author} {\bibfnamefont {N.}~\bibnamefont
  {Van~Thu}},\ }\bibfield  {title} {\bibinfo {title} {The casimir effect in
  bose–einstein condensate mixtures confined by a parallel plate geometry in
  the improved hartree–fock approximation},\ }\href
  {https://doi.org/10.1134/s1063776122080131} {\bibfield  {journal} {\bibinfo
  {journal} {Journal of Experimental and Theoretical Physics}\ }\textbf
  {\bibinfo {volume} {135}},\ \bibinfo {pages} {147} (\bibinfo {year}
  {2022})}\BibitemShut {NoStop}%
\bibitem [{\citenamefont {Salasnich}\ and\ \citenamefont
  {Toigo}(2016)}]{Salasnich2016}%
  \BibitemOpen
  \bibfield  {author} {\bibinfo {author} {\bibfnamefont {L.}~\bibnamefont
  {Salasnich}}\ and\ \bibinfo {author} {\bibfnamefont {F.}~\bibnamefont
  {Toigo}},\ }\bibfield  {title} {\bibinfo {title} {Zero-point energy of
  ultracold atoms},\ }\href {https://doi.org/10.1016/j.physrep.2016.06.003}
  {\bibfield  {journal} {\bibinfo  {journal} {Physics Reports}\ }\textbf
  {\bibinfo {volume} {640}},\ \bibinfo {pages} {1} (\bibinfo {year}
  {2016})}\BibitemShut {NoStop}%
\bibitem [{\citenamefont {’t Hooft}\ and\ \citenamefont
  {Veltman}(1972)}]{Hooft1972}%
  \BibitemOpen
  \bibfield  {author} {\bibinfo {author} {\bibfnamefont {G.}~\bibnamefont {’t
  Hooft}}\ and\ \bibinfo {author} {\bibfnamefont {M.}~\bibnamefont {Veltman}},\
  }\bibfield  {title} {\bibinfo {title} {Regularization and renormalization of
  gauge fields},\ }\href {https://doi.org/10.1016/0550-3213(72)90279-9}
  {\bibfield  {journal} {\bibinfo  {journal} {Nuclear Physics B}\ }\textbf
  {\bibinfo {volume} {44}},\ \bibinfo {pages} {189} (\bibinfo {year}
  {1972})}\BibitemShut {NoStop}%
\bibitem [{\citenamefont {Schakel}(2008)}]{Schakel2008}%
  \BibitemOpen
  \bibfield  {author} {\bibinfo {author} {\bibfnamefont {A.~M.}\ \bibnamefont
  {Schakel}},\ }\href@noop {} {\emph {\bibinfo {title} {Boulevard of broken
  symmetries}}}\ (\bibinfo  {publisher} {World Scientific},\ \bibinfo {address}
  {Singapore [u.a.]},\ \bibinfo {year} {2008})\ \bibinfo {note} {literaturverz.
  S. 357 - 372}\BibitemShut {NoStop}%
\bibitem [{\citenamefont {Lee}\ and\ \citenamefont {Yang}(1960)}]{Lee1960}%
  \BibitemOpen
  \bibfield  {author} {\bibinfo {author} {\bibfnamefont {T.~D.}\ \bibnamefont
  {Lee}}\ and\ \bibinfo {author} {\bibfnamefont {C.~N.}\ \bibnamefont {Yang}},\
  }\bibfield  {title} {\bibinfo {title} {Many-body problem in quantum
  statistical mechanics. iii. zero-temperature limit for dilute hard spheres},\
  }\href {https://doi.org/10.1103/physrev.117.12} {\bibfield  {journal}
  {\bibinfo  {journal} {Physical Review}\ }\textbf {\bibinfo {volume} {117}},\
  \bibinfo {pages} {12} (\bibinfo {year} {1960})}\BibitemShut {NoStop}%
\bibitem [{\citenamefont {Lee}\ \emph {et~al.}(1957)\citenamefont {Lee},
  \citenamefont {Huang},\ and\ \citenamefont {Yang}}]{Lee1957}%
  \BibitemOpen
  \bibfield  {author} {\bibinfo {author} {\bibfnamefont {T.~D.}\ \bibnamefont
  {Lee}}, \bibinfo {author} {\bibfnamefont {K.}~\bibnamefont {Huang}},\ and\
  \bibinfo {author} {\bibfnamefont {C.~N.}\ \bibnamefont {Yang}},\ }\bibfield
  {title} {\bibinfo {title} {Eigenvalues and eigenfunctions of a bose system of
  hard spheres and its low-temperature properties},\ }\href
  {https://doi.org/10.1103/physrev.106.1135} {\bibfield  {journal} {\bibinfo
  {journal} {Physical Review}\ }\textbf {\bibinfo {volume} {106}},\ \bibinfo
  {pages} {1135} (\bibinfo {year} {1957})}\BibitemShut {NoStop}%
\bibitem [{\citenamefont {Navon}\ \emph {et~al.}(2011)\citenamefont {Navon},
  \citenamefont {Piatecki}, \citenamefont {Günter}, \citenamefont {Rem},
  \citenamefont {Nguyen}, \citenamefont {Chevy}, \citenamefont {Krauth},\ and\
  \citenamefont {Salomon}}]{Navon2011}%
  \BibitemOpen
  \bibfield  {author} {\bibinfo {author} {\bibfnamefont {N.}~\bibnamefont
  {Navon}}, \bibinfo {author} {\bibfnamefont {S.}~\bibnamefont {Piatecki}},
  \bibinfo {author} {\bibfnamefont {K.}~\bibnamefont {Günter}}, \bibinfo
  {author} {\bibfnamefont {B.}~\bibnamefont {Rem}}, \bibinfo {author}
  {\bibfnamefont {T.~C.}\ \bibnamefont {Nguyen}}, \bibinfo {author}
  {\bibfnamefont {F.}~\bibnamefont {Chevy}}, \bibinfo {author} {\bibfnamefont
  {W.}~\bibnamefont {Krauth}},\ and\ \bibinfo {author} {\bibfnamefont
  {C.}~\bibnamefont {Salomon}},\ }\bibfield  {title} {\bibinfo {title}
  {Dynamics and thermodynamics of the low-temperature strongly interacting bose
  gas},\ }\href {https://doi.org/10.1103/physrevlett.107.135301} {\bibfield
  {journal} {\bibinfo  {journal} {Physical Review Letters}\ }\textbf {\bibinfo
  {volume} {107}},\ \bibinfo {pages} {135301} (\bibinfo {year}
  {2011})}\BibitemShut {NoStop}%
\bibitem [{\citenamefont {Mordini}\ \emph {et~al.}(2020)\citenamefont
  {Mordini}, \citenamefont {Trypogeorgos}, \citenamefont {Farolfi},
  \citenamefont {Wolswijk}, \citenamefont {Stringari}, \citenamefont
  {Lamporesi},\ and\ \citenamefont {Ferrari}}]{Mordini2020}%
  \BibitemOpen
  \bibfield  {author} {\bibinfo {author} {\bibfnamefont {C.}~\bibnamefont
  {Mordini}}, \bibinfo {author} {\bibfnamefont {D.}~\bibnamefont
  {Trypogeorgos}}, \bibinfo {author} {\bibfnamefont {A.}~\bibnamefont
  {Farolfi}}, \bibinfo {author} {\bibfnamefont {L.}~\bibnamefont {Wolswijk}},
  \bibinfo {author} {\bibfnamefont {S.}~\bibnamefont {Stringari}}, \bibinfo
  {author} {\bibfnamefont {G.}~\bibnamefont {Lamporesi}},\ and\ \bibinfo
  {author} {\bibfnamefont {G.}~\bibnamefont {Ferrari}},\ }\bibfield  {title}
  {\bibinfo {title} {Measurement of the canonical equation of state of a weakly
  interacting 3d bose gas},\ }\href
  {https://doi.org/10.1103/physrevlett.125.150404} {\bibfield  {journal}
  {\bibinfo  {journal} {Physical Review Letters}\ }\textbf {\bibinfo {volume}
  {125}},\ \bibinfo {pages} {150404} (\bibinfo {year} {2020})}\BibitemShut
  {NoStop}%
\bibitem [{\citenamefont {Spada}\ \emph {et~al.}(2022)\citenamefont {Spada},
  \citenamefont {Pilati},\ and\ \citenamefont {Giorgini}}]{Spada2022}%
  \BibitemOpen
  \bibfield  {author} {\bibinfo {author} {\bibfnamefont {G.}~\bibnamefont
  {Spada}}, \bibinfo {author} {\bibfnamefont {S.}~\bibnamefont {Pilati}},\ and\
  \bibinfo {author} {\bibfnamefont {S.}~\bibnamefont {Giorgini}},\ }\bibfield
  {title} {\bibinfo {title} {Thermodynamics of a dilute bose gas: A
  path-integral monte carlo study},\ }\href
  {https://doi.org/10.1103/physreva.105.013325} {\bibfield  {journal} {\bibinfo
   {journal} {Physical Review A}\ }\textbf {\bibinfo {volume} {105}},\ \bibinfo
  {pages} {013325} (\bibinfo {year} {2022})}\BibitemShut {NoStop}%
\end{thebibliography}%

\end{document}